\documentclass{aa}  

\usepackage{graphicx}
\usepackage{txfonts}
\usepackage{lipsum}
\usepackage{subcaption}         
\usepackage{placeins}         
\usepackage[hidelinks]{hyperref}

\usepackage{xcolor}

\begin{document}

   \title{Constraining the Galactic Center Dark Cluster with ELT/MICADO Observations}


   \author{Sebastiano D. von Fellenberg\inst{1,2}\fnmsep\thanks{Humboldt Feodor Lynen Fellow} 
   \and 
   Mat\'{u}\v{s} Labaj\inst{3}
   \and 
   Sean M. Ressler\inst{1}
   }

   \institute{Canadian Institute for Theoretical Astrophysics, University of Toronto, 60 St. George Street, Toronto, ON M5S 3H8, Canada \\
            \and Max Planck Institute for Radioastronomy, auf dem H{\"u}gel 69, Bonn, Germany \\
            \and Department of Theoretical Physics and Astrophysics, Masaryk University, Kotlářska 2, 611 37 Brno, Czech Republic \\}

   \date{Received September 30, 20XX}

 
  \abstract{The Galactic Center hosts the densest known stellar environment in the Milky Way, dominated by the massive black hole Sgr~A* and the surrounding nuclear star cluster. Theory predicts that this region should also contain a large population of stellar compact objects (SCOs)---black holes, neutron stars, and white dwarfs---forming a ``dark cluster'' whose distribution and properties remain observationally unconstrained. These unseen stellar remnants are central to questions of mass segregation, cluster dynamics, and the expected rate of extreme mass ratio inspirals (EMRIs) detectable by future gravitational-wave observatories including LISA. Current evidence for SCOs in the Galactic Center is indirect, relying on dynamical mass measurements, X-ray surveys, and a small number of transient sources. Direct detections remain elusive due to crowding, extinction, and the sensitivity limits of existing instruments.
We explore how upcoming facilities, in particular the Extremely Large Telescope (ELT) with its first-light imager MICADO, can fundamentally transform this field. MICADO’s combination of deep photometry, high spatial resolution, and precise astrometry will enable systematic searches for SCO-star binaries via photometric variability and orbital astrometric signatures, as well as direct detection of isolated accreting black holes interacting with the gas-rich Galactic Center environment. We outline the observational pathways, technical challenges, and expected sensitivities, showing that ELT/MICADO observations can provide the first quantitative constraints on the dark cluster population. Establishing these constraints will be pivotal for understanding the dynamical evolution of the Galactic Center, the role of compact remnants in nuclear star clusters, and the astrophysical context of gravitational-wave sources in galactic nuclei.
}

   \keywords{Galactic Center -- Black Holes -- Nuclear Star Clusters -- Gravitational Waves}

   \maketitle

\section{Introduction}


The Milky Way's Galactic Center (GC) harbors the densest known star cluster, the nuclear star cluster (NSC). Since the early 1990s, modern telescopes have been able to track stellar motions in this region, which has revealed a concentration of a large, unseen mass of $\sim 4\times 10^6~\mathrm{M_\odot}$. This was ultimately proven to be a massive black hole (MBH) associated with the radio source Sgr~A* \citep{Genzel1997,Ghez2003}. In the innermost region  \citep[$r < 1''$, e.g.,][]{Ghez2008, Gillessen2009}, stars effectively behave as test particles, with their dynamics primarily governed by the potential of the MBH. Tracking their motion enables a precise determination of the black hole’s gravitational potential and allows one to test general relativistic effects on their orbits \citep{GRAVITYCollaboration2018_redshift,GRAVITYCollaboration_schwarzschild, Do2019}.
At larger distances \citep[$r > 2''.5$, e.g.,][]{GravityCollaboartion2022_massdistribution}, the mass of the NSC itself becomes relevant, and statistical tests can be used to infer the presence of an additional enclosed mass of $\approx 6\times 10^6~\mathrm{M_\odot}$ \citep[e.g.,][]{Fritz2016}. 

Currently, only the most luminous stars are detectable because the light emitted from the Galactic Center (GC) is highly extinguished by dust along the line of sight \citep[e.g.,][]{Fritz2011}. Nevertheless, about $10^4$ stars have been individually identified, and their presence implies that many stellar compact remnants (SCOs---stellar  Compact Objects), such as black holes (BH), neutron stars (NS), or white dwarfs (WD), likely exist in the direct vicinity of the MBH \citep[e.g.,][]{MiraldaEscude2000,Freitag2006,Schoedel2020}. Because these objects have so far largely eluded detection, their presence, distribution, and properties can only be estimated by theoretical models. 

The question of how stars would settle around an MBH was first addressed by \cite{Peebles1972}, who argued that a steady state solution should exist in which there is a constant flux of stars through energy space, controlled by two-body interactions with a relaxation timescale, $\tau_{\rm{TB}}$. Building on this work, \cite{Bahcall1976} found that (equal-mass) stars settle into a power-law density profile, a so-called Bahcall---Wolf cusp:
\begin{equation}
    n(r) \propto r^{-7/4},
\end{equation}
where $n(r)$ is the radial number density and $r$ is the radial distance to the central mass.
\cite{Bahcall1977} further extended this model to account for a multi-mass distribution and showed that the different scattering efficiencies lead to a phenomenon called mass segregation, where different masses settle into different power-law slopes. The heavier particles sink inward, while lighter particles get pushed to larger radii. This general result has been confirmed and quantified by subsequent analytical arguments and N-body simulations \citep[e.g.,][]{Freitag2006,Alexander2009,Linial2022,Zhang2024}, which found that lighter stars should follow $n(r) \propto r^{-1.5}$, while the heavier components (such as stellar black holes) follow $n(r) \propto r^{-2}$.

Observationally, there is no direct confirmation of this otherwise clean theoretical prediction. The best stellar system at hand for direct tests, the GC, holds some surprises. 
\begin{itemize}
    \item The bright objects in the GC nuclear cluster do not comprise an old, relaxed system. Contrary to all predictions \citep[e.g.,][]{Morris1993}, the GC is an efficient star-forming region \citep{Krabbe1991, Paumard2006, Yelda2014}. More than 200 recently formed  Wolf-Rayet, O-, and B-type stars have been identified, with ages $\sim6~\mathrm{Myr}$ and a top-heavy initial mass function \citep[e.g.,][]{vonFellenberg2022,Jia2023}. These stars are found in disk-like structures, likely reflecting gaseous structures from which they were formed and they are clearly dynamically unrelaxed.
    \item Further in, where the orbital periods are short enough for full orbital solutions to be determined \citep[e.g.,][]{Gillessen2009,Boehle2016}, the cluster of young stars also defies all theoretical expectations: it is in a relaxed and isotropic distribution, despite the fact that the stellar ages are much shorter than the local two-body relaxation time. 
    \item Finally, the stars that come closest to the black hole (i.e., that have the smallest peri-center distance, $R_{\rm{p}}$) are lighter than the stars at a larger radial distance \citep{Burkert2024}, again in stark contrast to the theoretical expectation of mass segregation. 
\end{itemize}

\noindent These are long-standing riddles that have sparked interest in a broad community \citep[e.g.,][]{Alexander2017A}.
One possibility is that the observed stellar population may well be a poor tracer for the governing bodies of the underlying dark cluster. 
Regardless, a key problem in modern astronomy is determining the rate at which SCOs merge with MBHs in galactic nuclei throughout the universe.
Studying the gravitational waves emitted by  these SCO-MBH mergers, called extreme-mass-ratio inspirals (EMRIs), are a key science goal of the Laser Interferometer Space Antenna (LISA; \citealt{LISA_paper2017}). Because the distribution of SCOs is only (poorly) known theoretically, while unlikely to be zero, the number of EMRIs that LISA will observe is highly uncertain, with plausible estimates ranging from a few tens to several tens of thousands \citep{Amaro-Seoane2007,Amaro-Seoane2012}.  A direct measurement of the SCO distribution in the GC would thus be the Rosetta stone for understanding the nuclear star cluster dynamics. 

The Milky Way's GC is a unique laboratory to study these phenomena, as it is the only galactic nucleus close enough to obtain direct quantitative constraints on the so far unseen objects orbiting around the MBH Sgr~A*. This paper explores avenues toward detecting the dark cluster of SCOs in the GC, specifically those enabled by MICADO at the Extremely Large Telescope (ELT).

\section{Current constraints on the Galactic Center dark cluster}
\subsection{Extended, dynamical mass}
Observationally, the properties of the dark cluster are highly uncertain. The best constraints exist for the total mass contained in the Galactic Center, derived from the measurement of position, proper motions, and (in some cases) radial velocities of thousands of luminous stars. These measurements indicate that the NSC can be modeled as a so-called ``isotropic rotator'': stars move in random orientations but on average have the same sense of rotation as the Milky Way. 
Specifically, the total mass in the central $100$ arcseconds  of the galaxy is constrained to be $\sim 6\times 10^6~\mathrm{M_\odot}$ \citep[e.g.,][]{Chatzopoulos2015,Fritz2016,FeldmeierKrause2025}. 

At radial separations of a few arcseconds from Sgr~A*, stellar orbital periods drop below $100$ years, enabling precise determination of individual orbital elements. This is most strikingly exemplified by the star S2, the key tracer of Sgr~A*'s mass \citep[e.g.,][]{Ghez1998,Schodel2002,Gillessen2009,Gillessen2017,GRAVITYCollaboration2019_geometricDistance}. For stars with sufficiently short orbital periods, orbital solutions can simultaneously constrain both the mass of Sgr~A* and the surrounding extended mass distribution.
For instance, observations with GRAVITY have determined that the extended mass is $> 3000~\mathrm{M_\odot}$ within the orbit of the star S2 ($r \approx 0''.23$), and $\sim15000~\mathrm{M_\odot}$ within $r\approx3''$ of Sgr~A* \citep{GravityCollaboartion2022_massdistribution, GravityCollaboration2024_extendedMass}. 

\subsection{Stellar density profile}
In addition to the dynamical information, the stellar density profile (i.e., the number of stars per area) can be measured and compared to the theoretically expected ones. For a long time, it was unclear whether the Galactic Center's old giant star population indeed followed the expected relaxed  Bahcall--Wolf cusp \citep[e.g.,][]{Genzel2003_stars,Schoedel2007,Buchholz2009}. More recent studies seem to indicate that, while brighter giant stars do not appear to follow the expected distribution, at least the fainter population does appear to be relaxed \citep[$K_{\rm{mag}}=16 -20$, e.g.,][]{Schoedel2018,Habibi2019,Schoedel2020}. 

\subsection{Star formation history}\label{sec:sfhistory}
The so-called K-band luminosity function (KLF) relates the number of stars to their luminosity and it encodes the formation history of the giant star population ($M_{\rm{giant}} = 1 - 2 ~\mathrm{M_\odot}$).
Models of the KLF indicate that $80\%$ of the Galactic Center giant stars formed along with the galaxy some $\sim13~\mathrm{Gyr}$ ago. Nevertheless, subsequent star formation events occurred and contributed significantly to the present population, with major formation periods occurring $\approx 3- 4~\mathrm{Gyr}$ and $0.2- 0.5~\mathrm{Gyr}$ ago \citep[e.g.,][]{Schoedel2020}, as evidenced by spectroscopic observations \citep{Pfuhl2011}. In addition, a small fraction of stars \citep[$\sim 10^4~\mathrm{M_\odot}$ ;][]{Yelda2014} formed $6~\mathrm{Myr}$ ago, i.e., the Young Star cluster mentioned previously. While the more recent work by \cite{Pfuhl2011} and \cite{Schoedel2020} are in broad strokes consistent with each other and with previous work \citep[e.g.,][]{Blum2003,Figer2004,Maness2007}, \cite{Chen2023} measure a substantially different star formation history: they report a primary star formation event occurring just $\sim 5 ~\mathrm{Gyr}$ ago which formed $95\%$ of the stellar mass. While the datasets of the different studies are comparable or complementary in depth and quality, \cite{Chen2023} argue that their more accurate metallicity modeling leads to the difference in the inferred cluster properties.
The discussion of these differences is beyond the scope of this paper, and we adopt, in the following, values by \cite{Schoedel2020} primarily because they lead to more conservative estimates about the detectability of the dark cluster (i.e., the \cite{Chen2023} cluster model predicts compact objects that are easier to detect; see \autoref{sec:gcmodel} for more details).

\subsection{Stellar binarity}
Binaries play an important role in the Galactic Center: they enable the deposition of stars close to Sgr~A* via the so-called Hills mechanism \citep{Hills1988, Generozov2020}, strongly influence stellar dynamics \citep[for a review, see][]{Alexander2017A}, and can provide luminous counterparts to otherwise undetectable SCOs.
To date, only four binaries have been identified in the Galactic Center. All were discovered through dedicated variability studies, some of which were aided by spectroscopic follow-ups. To date, the most comprehensive studies were presented by \citet{Gautam2019,Gautam2024}, which were sensitive to binary periods ranging from $\sim 1$ to $\sim 4000$ days, and included 563 stars, of which roughly half were found to be intrinsically variable. A subset of $12$ candidate binary stars was found ($2\%$ of the surveyed stars), three of which had variability on time scales of $P=1-10$ days, five with $P=1-80~\mathrm{d}$, and four with $P>80~\mathrm{d}$. 
The binary system IRS 16SW has a photometric period of 9.6 days, with a minimum total mass of $100~\mathrm{M_{\odot}}$ \citep{Ott1999new}, and an equal mass ratio for the components \citep[][though note that these authors did not discuss the possibility of a star-SCO binary]{Martins2006}. 
The second known binary is IRS 16NE, which was discovered via radial velocity measurements, a long-period eclipsing binary with a period of $\sim 200$ days, a combined mass of $\sim 30M_{\odot}$ and a mass ratio of $2\pm0.5$. 
The third binary, E60, was again discovered via photometry and is a close eclipsing binary with an orbital period of roughly two days \citep{Pfuhl2014}. 
The fourth binary, S2-36, which shows a period of $\sim 40$ days, was discovered photometrically \citep{Gautam2019} and confirmed by radial velocity measurements \citep{Gautam2024}. All of these binary systems are young stars, i.e., they were formed in the most recent star formation event $\sim 6~\mathrm{Myr}$ ago \cite[e.g.,][]{Paumard2006, Yelda2014,vonFellenberg2022}.
In addition to those candidate systems, a few more candidate systems have been proposed, but have so far not been robustly confirmed: IRS29N shows features associated with a wind colliding binary \citep{Rafelski2007}; based on H-R diagram modeling \cite{Geballe2006} suggest IRS8 might be a binary; and IRS7E shows a large variability in radial velocity potentially consistent with a binary \citep{Paumard2006}. Most recently, \cite{Peissker2024_binary} reported a binary system in a Galactic Center G-object.

\subsection{Direct evidence for Galactic Center compact objects}
So far, three bona fide direct measurements of SCOs in the Galactic Center have been reported. The first, XJ174540.0–290031, is a radio/X-ray transient that peaked at radio frequencies in 2004 \citep{Bower2005, Porquet2005}, located $2.9\mathrm{''}$ from Sgr~A*. Although its X-ray emission faded, it remains a compact radio source \citep{Zhao2022}, likely a quiescent low mass X-ray binary (qLMXB) or a free-floating SBH that underwent a strong accretion event. Notably, it co-moves with the northern arm of the mini-spiral, potentially tapping into an additional gas reservoir \citep{Bower2005}. The second source, Sgr~J1745–2900, is a magnetar located $\sim 2.4~\mathrm{''}$ from Sgr~A*, first detected in X-rays \citep{Eatough2013, Kennea2013} and later in radio \citep{Bower2014}. Its proper motion aligns with the GC clockwise disk of young stars \citep{Bower2015}, suggesting it originated from the most recent star formation episode $\sim 6~\mathrm{Myr}$ ago. The third source, MAXI J1744–294, was discovered in early 2025 as an X-ray transient \citep[e.g.,][]{Kudo2025ATel} and subsequently in radio \citep[e.g.,][]{Michail2025ATel}. Follow-up observations identify it as an LMXB \citep[][]{Mandel2025arXiv}, and this conclusion has been consistently confirmed across multiple telescopes and groups \citep[e.g.,][]{Marra2025arxive, Chatterjee2025}.

Beyond these three bona fide SCOs, several candidate sources have been reported. \citet{Zhao2022} conducted a deep JVLA survey of the central few arcseconds and identified dozens of compact radio sources. While some were linked to stellar winds based on their SED, a significant fraction could be consistent with radio emission from SCOs (i.e., SBHs, NSs). An extended X-ray source (G359.95-0.04), typically interpreted as a pulsar wind nebula, has also been detected \citep[e.g.,][]{Zhao2022}. Further evidence for SCO-star binaries comes from X-ray surveys with Chandra and XMM-Newton, which have revealed faint point sources through Fe XXV $K\alpha$ line emission \citep{Wang2002}. \citet{Muno2003,Muno2005,Muno2009} and \citet{Zhu2018} cataloged approximately $\sim 3500$ such sources, which roughly trace the stellar light distribution; however, source confusion within $5''$ of Sgr~A* limits the sensitivity to faint objects. \citep{Muno2005} suggest an overabundance of such systems in the central parsec compared to the region outside of the NSC, in line with expectation from mass segregation \citep{Bahcall1977}.
These point sources are generally considered X-ray counterparts to SCO-star binaries, and their population is dominated by magnetic Cataclysmic variables (MCVs), i.e., white dwarf–star systems. 
qLMXBs appear subdominant ($\sim 2\%$), consistent with \citet{Hailey2018}, who reported only $12$ SBH containing qLMXBs in the Chandra dataset, most centrally concentrated. Consistent with this conclusion, \cite{Krivonos2021} suggested that out of $20$ detected GC very faint X-ray transients, $12$ likely correspond to qLMXBs.

\subsection{Galactic Center pulsars}
Pulsars, a distinct class of SCOs, are short-lived, highly magnetized neutron stars emitting periodic radio flashes. 
From theoretical arguments, there is expected to be $N_{\rm{pulsar}} \sim \mathcal{O}(0.1-10)$ pulsars in the GC \citep{Eatough2013,Schoedel2020}, but so far none have been detected. 
Such pulsars, if they exist, could probe the potential of Sgr~A* with high precision \citep[e.g.,][]{Kramer2004,Cordes2004,Pfahl2004}.
Their detection requires deep, high-cadence radio/mm surveys, which are ongoing with Meerkat, ALMA, and soon SKA \citep[e.g.,][]{Kramer2013,Kramer2015,Eatough2015,Mus2022,Frail2024}. Given extensive prior work, pulsars are excluded from this paper’s scope.  

\section{Current detection strategies for compact objects in the optical and infrared bands}
Four general pathways for the detection of compact objects exist:
\begin{enumerate}
    \item Photometric searches,
    \item Spectroscopic searches,
    \item Astrometric searches,
    \item Gravitational lensing searches.
\end{enumerate}
We discuss each of these below.

\subsection{Photometric searches}
Photometric searches for compact objects generally rely on detecting periodic variability in the light curves of gravitationally bound binary systems. Such variability is caused by the gravitational influence of the companion object, which deforms the primary, modulates its luminosity, and thereby imprints the orbital dynamics on the light curve \citep[e.g.,][]{Morris1985}. The strength of the imprinted variability depends on the binary mass ratio, with higher variability amplitude in tighter systems and greater mass ratios \citep[][]{Morris1993_elips}.
This variability can be significant, reaching $\sim20\%$ for heavy, tightly bound systems \citep[][]{Gomel2021}.   
However, only the mass ratio, and not the two individual masses, can be derived with photometry. Because ellipsoidal variability arises in both star–star binaries and star–compact object systems \citep[e.g.,][]{Moe2017,Gomel2021_mbmr}, in order to constrain the nature of the unseen secondary object the mass of its luminous companion must be known, but is often highly uncertain.
Photometric searches have been greatly enhanced by large catalogs constructed by space observatories such as TESS \citep{Ricker2015_tess}, OGEL \citep{Udalski2015_ogle}, and Gaia \citep[][]{Gaia2016}, which survey millions of stars for periodic features in their light curves. 
For instance, \cite{Gomel2021} presented a catalog of 10956 short-period ($P<2.5~\mathrm{d}$) ellipsoidal binaries, out of which they identified $136$ candidate star-SCO binary systems.
\subsection{Spectroscopic searches}
Spectroscopic searches of SCO-star binary systems are, in principle, similar to photometric searches, but probe periodic variations in the radial velocity of the luminous counterpart instead of its light curve. As in photometric surveys, SCO-star binaries need to be disentangled from the population of common star-star binaries, and are often identified as a by-catch of binary star surveys \citep[e.g.,][]{Giesers2018,Shenar2022,Mahy2022}.
Such surveys have detected credible candidate systems in massive, single-lined (SB1) or double-lined (SB2) O/B-type binary systems, with typical orbital periods of a few tens of days. For instance, \cite{Shenar2022} characterized $51$ candidate SCO-O/B-star binary systems, out of which they identified one robust O+BH binary system \citep{Shenar2022_BHdetc}, and two additional high-confidence candidate systems.
Although spectroscopic searches provide strong constraints on system masses, the necessity of dispersing the light significantly reduces their efficiency. 

\subsection{Astrometric searches}
Astrometric searches combine the advantages of photometric and spectroscopic approaches, by using photometry as an efficient survey technique and velocity measurements to derive accurate binary system mass measurements. 
Specifically, the high astrometric accuracy of the Gaia mission \citep{Gaia2016} makes it a promising instrument to identify Star-SCO binary systems \citep[e.g.,][]{Breivik2017,Mashian2017,ElBadry2021_gaiacat}. Although the number of potentially detectable sources is uncertain, high confidence SCO-star binaries have already been detected with Gaia \citep{ElBadry2023_redgiant,ElBadry2023_sunlike, ElBadry2024_nuetronstars}.

\subsection{Gravitational lensing searches}
Finally, SCOs can be revealed through the gravitational magnification of a background source, a phenomenon known as microlensing. Originally proposed by \citet{Einstein1936}, microlensing enables serendipitous detections of isolated compact objects acting as lenses along with a precise determination of their mass. 
Modern, optimized surveys such as the Optical Gravitational Lensing Experiment \citep[OGLE;][]{Udalski2015_ogle}, Microlensing Observations in Astrophysics \citep[MAO;][]{Sumi2016MAO} or the Zwicky Transient Facility \citep[ZTF;][]{Masci2019ZTF} routinely detect microlensing events, and the technique has become a prime method for discovering exoplanets, with more than $200$ detections to date \citep[e.g.,][]{Mroz2023}. In principle, lenses of any mass can be detected with this method; however, to date, only a single isolated stellar-mass black hole has been confirmed \citep{Sahu2022}.

\section{Detection strategies for Galactic Center SCOs with the ELT}
\subsection{A simplified Galactic Center model}\label{sec:gcmodel}
\begin{figure}
    \centering
    \includegraphics[width=0.5\textwidth,trim=0 0 0 1.5cm,  clip]{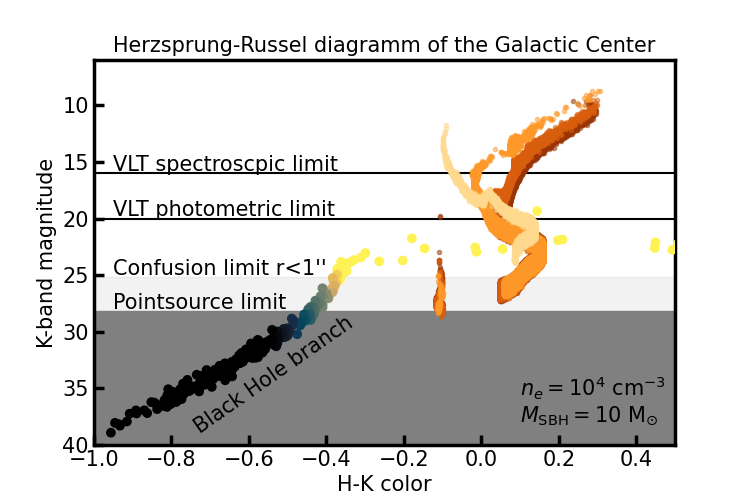}
    \caption{Model Hertzsprung---Russell diagram of the Galactic Center. Horizontal lines indicate sensitivity limits of the VLT as annotated, shaded regions correspond to the respective ELT/MICADO limits. The orange stellar evolutionary tracks are based on MIST isochrones \citep{Dotter2016}, and use the GC mass reported by \citet{Fritz2016}, as well as the age and star formation history reported by \citet{Schoedel2020}. The light orange track shows the Young Star cluster \citep{vonFellenberg2022}. For all stars, we assume a Kroupa IMF \citep{Kroupa2001}, which may represent a poor approximation in the Galactic Center \citep[e.g.,][]{Lu2013}. The annotated, black-to-yellow colored ``Black Hole branch'' shows the luminosity of modeled SBHs accreting from the Mini-spiral as discussed in \autoref{sec:directDetec}. An artificial scatter of a few percent has been added to the tracks for visualization purposes.}
    \label{fig:HR-diagram}
\end{figure}

Photometric, spectroscopic, and astrometric SCO searches generally rely on the presence of a luminous star in a bound star-SCO binary system, while microlensing only requires a luminous background source along the line of sight.
It is thus clear that, in order to estimate the detectability of SCOs in the Galactic Center, we need to estimate the luminosity and distribution of all Galactic Center components, i.e., we need to define a baseline model for the Galactic Center.
In order to obtain a realistic, albeit approximate distribution of stars, we model the three significant star formation events measured by \cite{Schoedel2020}. To do so, we use the stellar evolutionary tracks based on the MESA Isochrones \& Stellar Tracks (MIST) isochrones \citep{Dotter2016}. To estimate the number of stars, we adopt the Galactic Center mass from \citep{Fritz2016}, and we model all known major star formation events: $80\%$ of mass formed $\approx10 ~\mathrm{Gyr}$ ago, $15\%$ of mass formed $\approx3 ~\mathrm{Gyr}$ ago, and $5\%$ of mass formed $\approx250~\mathrm{Myr}$ ago \citep[based on][]{Schoedel2020}. In addition, we include the Young Star cluster assuming a cluster mass of $10^4~\mathrm{M_{\odot}}$ and an age of $6 ~\mathrm{Myr}$ \citep[e.g.,][]{Bartko2009,Lu2013}. For simplicity, we adopt a Kroupa Initial Mass Function \citep[IMF;][]{Kroupa2001} for the stars.
\autoref{fig:HR-diagram} shows the resulting Hertzsprung--Russell diagram. The distance modulus and extinction are accounted for in the stellar flux density (y-axis). 
While we could use the MIST isochrones to estimate the number of stars, white dwarfs (WD), neutron stars (NS), and SBHs, we adopt the numbers from \cite{Zhang2024} and \cite{GravityCollaboration2024_extendedMass} instead. These authors provide estimates based on much more sophisticated Fokker--Planck modeling and observational constraints of the NSC. For the following estimates, based on their modeling and observations, we assume that $10^7$ stars, $5\times10^5$ WDs, $10^5$ NS, and $10^4$ SBHs are present in the Galactic Center Nuclear Star Cluster (i.e., within $\approx 30''$ of Sgr A*).

Finally, the number of compact objects residing in binary systems remains poorly constrained. Although observations show that binaries are present—at least within the younger stellar populations of the NSC (see Introduction)—their overall abundance in the GC is governed by several competing processes: the primordial binary fraction, the efficiency with which star–star binaries capture SCOs, and the subsequent evaporation of binaries through dynamical encounters in the dense cluster environment \citep[for reviews, see][]{Alexander2005,Alexander2017A}.
Outside the Galactic Center, stars predominantly form in binaries, with the binary fraction increasing strongly with primary mass \citep[e.g.,][]{Sana2013,Sana2014}. Given the top-heavy stellar mass function in the GC \citep[$\alpha  \in (-0.45,+1.7)$; e.g.,][]{Bartko2010,Lu2013,Do2013,vonFellenberg2022}, one would therefore expect an even higher binary fraction than in the rest of the Milky Way. However, the unusual, disk-like mode of star formation in the GC \citep[e.g.,][]{vonFellenberg2022,Jia2023} cautions against a straightforward extrapolation of Galactic disk binarity to the NSC. Still, it seems plausible that most stars that ultimately form SBHs and NSs are formed in binary systems. 

The stellar density in the GC is too low for binaries to form efficiently through dynamical capture \citep[e.g.,][]{Pfuhl2014}. The exchange capture of a more massive component, such as a stellar-mass black hole, by a star-star binary via the Hills mechanism \citep[][]{Hills1988} is possible, but the rate of such events remains uncertain, as it depends sensitively on the (largely unknown) population of isolated SBHs in the region.

Still, the fact that binaries are observed in the Galactic Center \citep[e.g.,][and see Introduction]{Pfuhl2014} demonstrates that they do form and can survive for at least part of their evolutionary history. Once formed, binaries interact with both the surrounding NSC and the central SMBH. Repeated two-body encounters tend to drive their evolution toward either coalescence or evaporation \citep[e.g.,][]{Alexander2005}. Coalescence timescales for hard binaries are long \citep[comparable to the local relaxation time, i.e.,\ $\sim 10^{9\text{--}10}~\mathrm{yr}$, e.g.,][]{Merritt2010}. In contrast, evaporation timescales can be much shorter and depend sensitively on the distance from the SMBH, the binary semi-major axis, and the masses of both the binary components and the surrounding cluster stars \citep{Pfuhl2014,Alexander2014}. 

Massive binaries, the progenitors of SBHs, are generally stable over their stellar lifetimes. However, once the SBHs-star binary system forms, even relatively tight systems ($a_0 < 1~\mathrm{AU}$) are expected to evaporate over the lifetime of the NSC \citep[$\sim 10~\mathrm{Gyr}$, e.g.,][]{Pfuhl2014, Marklund2025}. Recent estimates by \cite{Dodici2026} suggest that all but the most tightly bound binaries [$a_0(r>1~\mathrm{pc}) < 1~\mathrm{AU}$ and $a_0(0.01~\mathrm{pc}<r<0.1~\mathrm{pc}) < 0.01~\mathrm{AU}$] will evaporate within $10^{10}~\mathrm{yr}$. This implies that the oldest and most numerous stellar populations in the NSC ($\sim 80\%$) are unlikely to retain a significant binary fraction. Nevertheless, a non-negligible fraction of binaries may persist among the more recently formed stars ($\sim 20\%$). 

Finally, SBHs can grow substantially in mass via mergers \citep[e.g.,][]{Doctor2020,LVK_bbmassspec2025arXiv}. NSCs are of particular interest in this context as the dense stellar fields allow for multiple pathways for SCOs to merge and form heavier merger products \citep[e.g.,][]{Hoang2018, Kremer2020}. For instance, by explicitly modeling the merger product population in the Galactic Center, \cite{Newton2026} find that SBHs can reach masses of up to $\sim 200~\mathrm{M_{\odot}}$ and predict the presence of several tens of such objects. They estimate that although more massive products may form, they efficiently sink and merge with Sgr A* and are therefore only transiently present.

Given these uncertainties and the simplified NSC model adopted here, we do not attempt to derive a detailed or self-consistent binary fraction. Instead, we account only for the plausible presence of binaries. As discussed in \autoref{sec:sfhistory}, we note that \cite{Chen2023} derive a substantially different star formation history, and importantly derive an estimate of about an order of magnitude larger for the number of SBHs present in the Galactic Center ($2.5\times 10^5$). They also model the SBH-SBH binary fraction explicitly, and estimate the presence of $2.2\times 10^4$ SBH-SBH binaries, but do not report a SCO-star-binary fraction. We show in what follows that under any circumstances only a small fraction of the SCOs are likely detectable in the Galactic Center, thus such an increased SBHs population would trivially correspond with a larger number of detectable systems.

\subsection{ELT/MICADO observations of the Galactic Center}
The Extremely Large Telescope (ELT) is the flagship ground-based optical and near-infrared observatory of the European Southern Observatory (ESO), currently under construction on Cerro Armazones in northern Chile \citep{ESO_ELT}. With a 39~m diameter segmented primary mirror, the ELT will provide an unprecedented combination of light-gathering power and angular resolution, enabling advances across stellar astrophysics, galaxy formation, and fundamental physics \citep[e.g.,][]{Gilmozzi2007}. The telescope is designed to operate in conjunction with advanced adaptive optics systems, allowing it to achieve diffraction-limited performance at near-infrared wavelengths \citep[e.g.,][]{Bonnet2018}.
One of the ELT’s first-light instruments is MICADO (Multi-Adaptive Optics Imaging Camera for Deep Observations), a high-resolution near-infrared imager optimized for precision astrometry and deep imaging \citep{Davies2010,Davies2021,Sturm2024}. At first, MICADO will operate using a single-conjugate natural guide star adaptive-optics system (SCAO) system developed by the MICADO and MAORY consortia, and will subsequently be upgraded to operate behind the multi-conjugate adaptive optics module MAORY, which delivers uniform, high-Strehl imaging over fields of view up to $\sim 1'$ \citep{Diolaiti2016maory}.

MICADO observations of the Galactic Center will be transformative because they overcome or greatly reduce the two main obstacles limiting 10-m-class telescope observations. 
First, the high spatial resolution of $\sim 10~\mathrm{mas}$ in the K-band greatly reduces confusion in the Galactic Center. While the stellar density in the central arcsecond of  $10^4~\mathrm{stars/as^2}$ still implies confusion-limited observations for the central few arcseconds, the surface density of stars drops off steeply with distance and thus outside of the central arcsecond the observations should be limited by integration time. 
This picture is complicated by the presence of several very bright stars, specifically IRS7 \citep[$K_{\rm{mag}}\approx7\mathrm{mag}$; e.g.,][]{Ott1999new} and the IRS16 sources \citep[$K_{\rm{mag}}\approx 10\dots12~\mathrm{mag}$; e.g.,][]{Martins2007} which will likely limit observations in their immediate spatial vicinity (which includes Sgr~A*). This may require specific mitigation strategies such as a field stop, and excellent AO-performances over a large field of view (FOV).
Assuming these technical limitations can be overcome, photometric sensitives should range from $K_{\rm{mag}}=25$ in the confusion-limited central arcsecond and approach $K_{\rm{mag}}=29$ at larger projected distances for a typical, July/August observation of the Galactic Center \citep[$t_{\rm{obs}}\approx 6~\mathrm{h},$][]{Davies2021}. These limits are indicated by light and dark gray shaded regions in \autoref{fig:HR-diagram}.
Finally, MICADO will have excellent relative astrometry, allowing for $\sim 50~\mu as$ precision  during a typical good observing night \citep{Davies2021}.

\subsection{Photometric detection of SCO-star systems in the Galactic Center}
\begin{figure}
    \centering
    \includegraphics[width=0.485\textwidth,trim=0cm 0cm 17.25cm 0cm, clip]{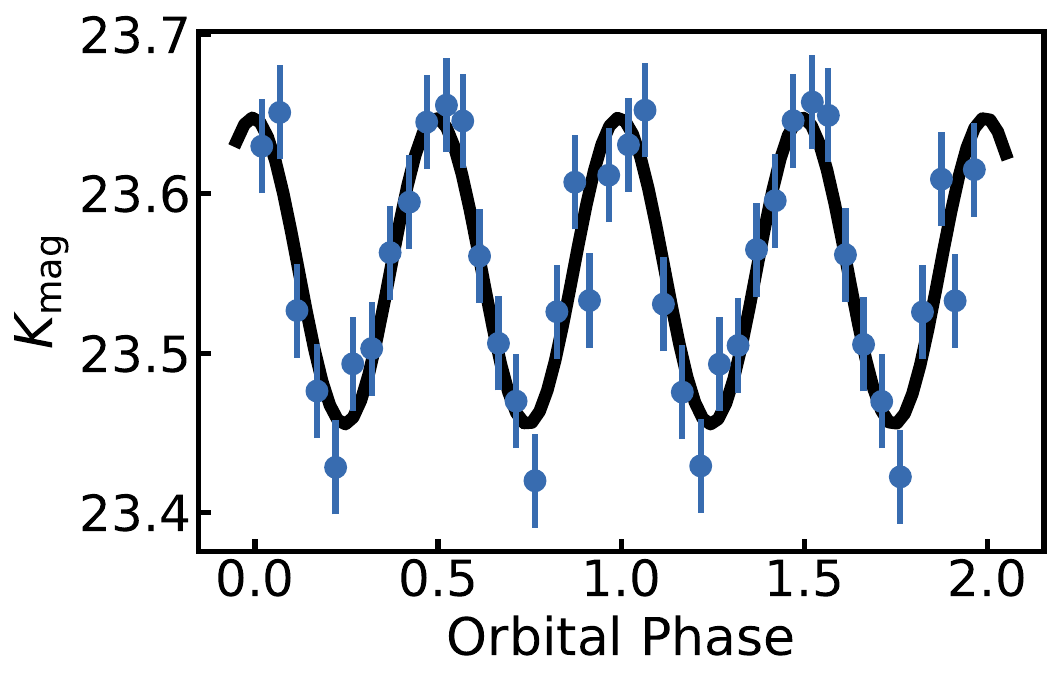}
    \caption{Mock observations of the variability of the host K7V main sequence star in the LMXB XTE J1118+480, scaled as if it were located in the Galactic Center. The star has a K-band magnitude $\sim 23.6$, which is readily detectable by the ELT \citep{Gelino2006}, assuming a constant extinction of $A_K=2.42$ for simplicity. There are differences in star flux density of a factor $\times10$ between references \citep[i.e.,][]{Plotkin2015}, but the system is readily detected in any case. We have assumed $10\%$ photometric uncertainty.}
    \label{fig:J1118}
\end{figure}

Photometric detection of SCOs, i.e., the identification of ellipsoidal binaries, requires the presence of a luminous companion star. Importantly, the ELT will allow for the detection of the G- and K-type stars that typically represent host stars in known Low-Mass X-ray Binaries (LMXBs), composed of a main-sequence star and an SCO.
To illustrate the detectability of such a system, it is useful to study a known LMXB in which the stellar system is well-characterized, and scale its luminosity to the Galactic Center distance. 
For instance, in the case of XTE J1118+480, the host star is a K7V main sequence star, with a K-band magnitude of $K_{\rm{mag}} \sim 16.7,\,\Delta K_{\rm{mag}} \sim 0.3$, and a photometric period of $P \sim53 ~\mathrm{d}$ \citep{Gelino2006}. 
If this system was placed in the Galactic Center, the host star would have a minimum magnitude of  $K_{\rm{mag}} \sim 23.6$,  see \autoref{fig:J1118}. 
Although the error bars were artificially imposed on the data (assuming $10\%$ photometric noise), it is evident that such a system would be readily detectable by the ELT within a year, provided the Galactic Center is monitored on a semi-regular basis ($\sim 1/\mathrm{month}$), as is routinely done today with the VLT. 
This discussion illustrates that, in principle, the detectability of such systems is not inherently difficult and requires only moderately regular cadence observations, comparable to those conducted in the Galactic Center over the past two decades. At the same time, even very conservative (i.e., $\sim 1\%$) estimates of the binary fraction in the Galactic Center predict the presence of $\sim10^5$ ordinary star–star binary systems, which constitute a significant astrophysical (rather than instrumental) noise source in the identification of SCO–star binaries. 

It is therefore evident that addressing this problem demands survey-scale analysis strategies that have not yet been systematically applied to the Galactic Center. Nevertheless, constraining the photometric binary population across the entire stellar content of the Galactic Center, independent of the nature of the companion object, represents an obvious and compelling research direction for the upcoming ELT era. The eventual detection of elusive star–SCO candidate systems may well be its most rewarding outcome.

\subsection{Spectroscopic detection of SCO-star systems in the Galactic Center}
MICADO is currently only envisioned to offer a slit spectrometer. Due to the extreme confusion in the Galactic Center, slit spectrometry is inefficient for survey style searches and can only be used to track a known, bright star (such as S2). Therefore, while individual promising and bright enough candidate star-SCO systems maybe followed spectroscopically, systematic spectroscopic searches will have to wait for Integral Field Unit (IFU) instruments such as HARMONI \citep{HARMONIESO2021} or ANDES \citep{AndesESO2024}, and we defer the details to future work. 

\subsection{Astrometric detection of SCO-star systems in the Galactic Center}
The last few years have illustrated that, thanks to the astrometric precision of the Gaia mission, astrometric detection of compact objects is both feasible and robust \citep[e.g.,][]{ElBadry2023_sunlike,ElBadry2023_redgiant,ElBadry2024_nuetronstars}. Gaia achieved astrometric accuracies $\sim 20~\mathrm{\mu as}$ in its third data release \citep{gaia2023_DR3}. The ELT/MICADO will achieve a comparable (relative) astrometric accuracy of $\sigma_{\rm{MICADO}}\approx 50~\mathrm{\mu as}$ \citep[][]{Massari2016,Davies2021}. At a distance of $8.2~\mathrm{kpc}$, this corresponds to a projected spatial scale of $\sim 0.4~\mathrm{AU}$. 

\begin{figure*}
    \centering
    \includegraphics[width=0.985\textwidth]{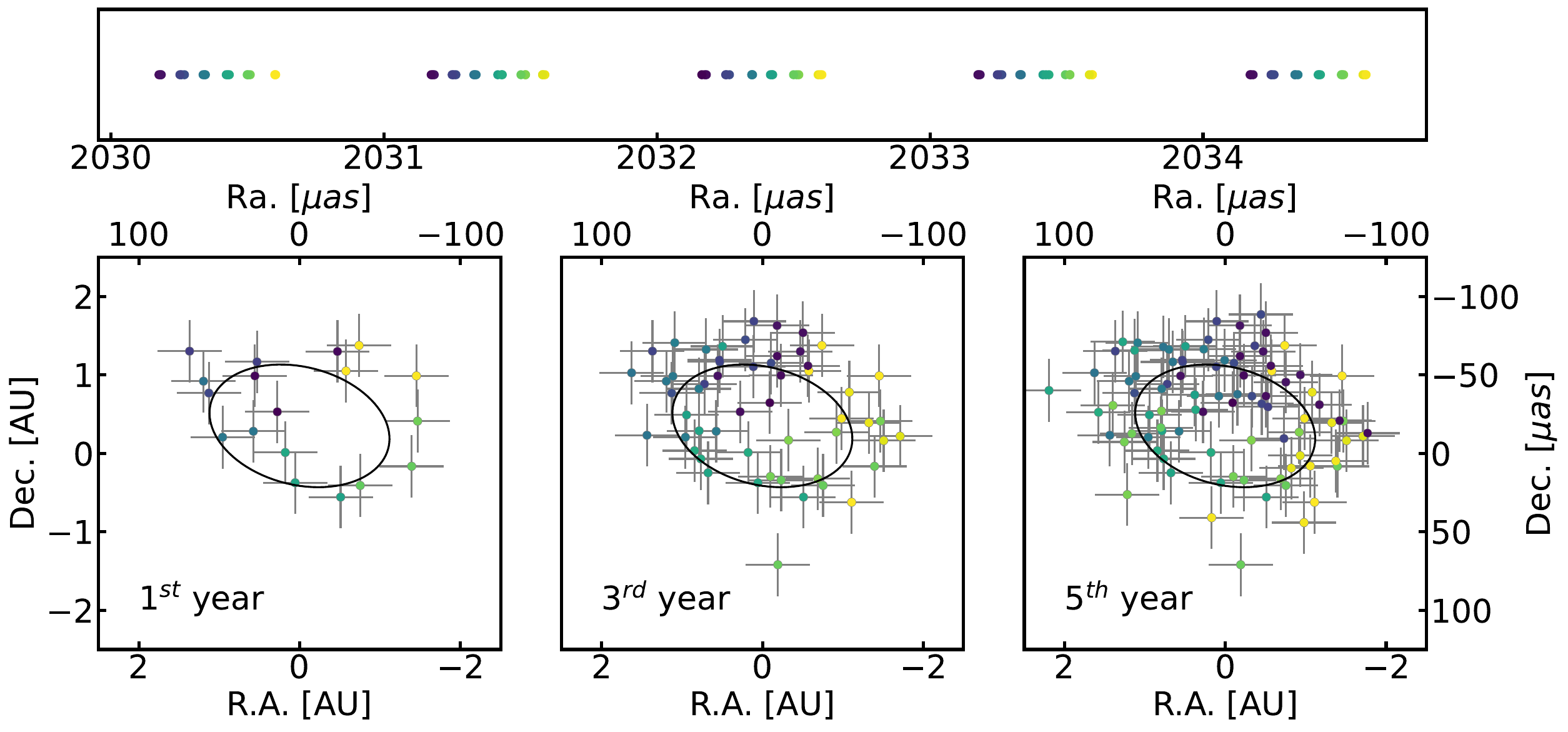}
    \caption{Mock astrometric observation of the Gaia-detected star-stellar mass black hole binary system \citep{ElBadry2023_sunlike} if it were located in the Galactic Center. We scale the orbital elements to the distance of the Galactic Center (black solid line), assume a regular observing cadence during the Galactic Center season (top panel, color indicates observing month), and an astrometric accuracy of $50\mathrm{\mu as}$. For simplicity, we assume perfect subtraction of the proper motion of the binary in the Galactic Center potential, and neglect astrometric confusion (which can occur if the system crosses paths with another star).}
    \label{fig:gaiaBH}
\end{figure*}

To demonstrate the feasibility of astrometric SCO-star detections, we use an observed example: a black hole discovered in a sun-like star system \citep{ElBadry2023_sunlike}. This system, discovered in the Gaia astrometry and radial velocity sample, consists of a G-type star primary and is found at distance of $ 480~\mathrm{pc}$. The hidden black hole companion has a mass of $9.62~\mathrm{M_\odot}$. Comparing with \autoref{fig:HR-diagram}, it is clear that the host star (flux density $\sim 22 ~\mathrm{mag}$) would be readily detectable if it were located in the Galactic Center. The system has a semi-major axis of $1.40\pm0.01~\mathrm{AU}$ and an orbital period of $P\approx 185~\mathrm{d}$. Using Thiele-Innes elements \citep[e.g.,][]{Breivik2017} and the orbital parameters reported in \cite{ElBadry2023_sunlike}, we project the system as if it were observed in the Galactic Center (\autoref{fig:gaiaBH}). We use a realistic observing cadence of three images per month over five years, and assume a nominal uncertainty of $50~\mathrm{\mu as}$. 
\autoref{fig:gaiaBH} shows that such a measurement is feasible, and robust measurements of the system's orbit would be available after a few years.

Next, we try to estimate the number of SCO-star binaries that could be detected with MICADO. For the purpose of this calculation, we assume a binary fraction of $50\%$, yielding 5000 binary systems based on the assumption of $10^4$ SBHs within $\approx 30''$. This assumption, while somewhat arbitrarily, is consistent with the high binary fraction of massive stars \citep[e.g.,][]{Sana2013}.
Additional binaries may be formed by tidal capture, but these are expected to be a small fraction of the population \citep{Generozov2018}. Because the total number of binaries in the GC is poorly constrained \citep[e.g.,][]{Gautam2019}, we do not attempt to obtain an absolute estimate of detectable systems; instead we derive a fractional estimate. The binary orbital properties are adopted from \cite{Dodici2026}, who demonstrated that a significant fraction of initially wide binaries evolve into near-contact systems through von Zeipel–Lidov–Kozai (ZLK) oscillations. They further report a radial dependence of the binary semi-major axis. For the purpose of this calculation, we use their semi-major axis distribution for two radial bins: $0.1~\mathrm{pc}$ to $1~\mathrm{pc}$ and $>1~\mathrm{pc}$. We further assume a thermal eccentricity distribution and a black hole mass of $10~\mathrm{M_{\odot }}$, while we adopt the stellar mass distribution from \autoref{sec:gcmodel} and use Thiele-Innes elements to calculate the projected orbits. Under these assumptions, we find that approximately $\sim 15\%$ of the SCO-star binaries exhibit orbital variations large enough to be detectable astrometrically at the $3\sigma \, =\, 1.2~\mathrm{AU}$ level. 
If instead we assume that the binary orbits are circular, this estimate is not significantly altered ($\sim 18\%$ detected).
Furthermore, we emphasize that our fractional estimate for detectability is independent of the assumed total number of systems; it depends only on their orbital properties.

\begin{figure}
    \centering
    \includegraphics[width=0.485\textwidth]{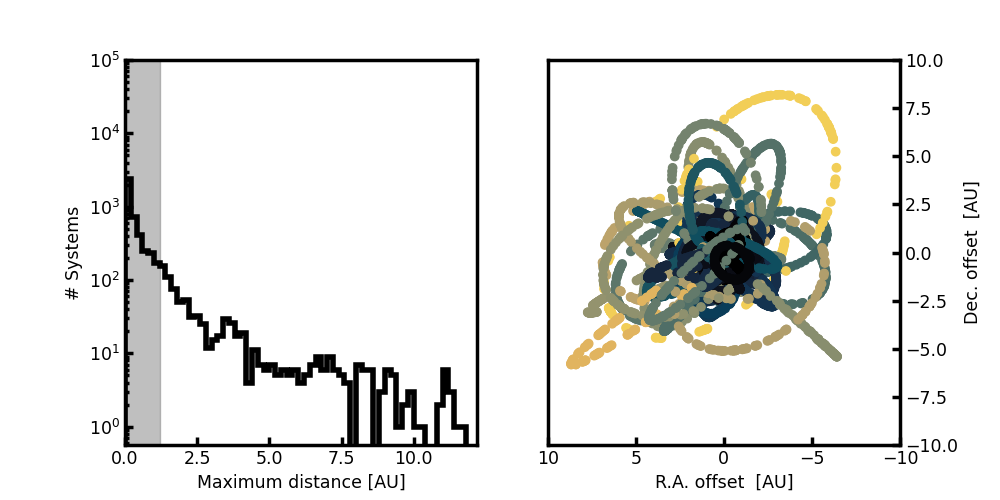}
    \caption{Left: Maximum size of orbital ellipses for simulated star-SCO binaries using the semi-major axis distributions in \cite{Dodici2026} as input. The gray shaded region shows the $\gtrsim  3\sigma$ detection limit assuming $50~\mathrm{\mu as}$ astrometric uncertainty. Right: $100$ sample orbits for binaries that would be detectable with high significance.}
    \label{fig:astroBinaries}
\end{figure}

\autoref{fig:astroBinaries} presents the resulting distributions for star-SCO binaries, along with a representative sample of orbits for which astrometric detections should be feasible.
Of course, this estimate is highly uncertain, as neither the fraction of stellar compact objects (SCOs) bound in binary systems nor their orbital properties are well constrained.
Specifically, the total number of detectable systems is a function of the binary fraction, and the percentage of detectable stars sensitively depends on the binary semi-major axes distribution. If the binary semi-major axis is too small, that system will not be detectable; the calculations by \cite{Dodici2026} indicate that this might be the case at radial separations from Sgr A* smaller than $\sim 0.1~\mathrm{pc}  \approx 2.5$ arcseconds, where the fraction of detectable binaries drops to $\sim 1\%$. On the other hand, \cite{Dodici2026} focus only on the main-sequence phase of stellar binaries and do not account for modifications occurring during later evolutionary stages. These later stages, such as the common-envelope phase, are expected to significantly alter the binary orbital parameters \citep[e.g.,][]{Soberman1997}.
If this radial dependence of the binary semi-major axis is correct, it would also mean that both photometric and astrometric methods have complementary sweet-spot regimes.  
Specifically, this occurs because closer to Sgr A*, near-contact binaries with strong photometric variability are more common, while at further distances larger semi-major axis systems are more easily detected via astrometric means.
In addition to the unkown properties of the SCO-star binaries, we neglect the uncertainty stemming from subtracting the intrinsic proper motion of the system, which, bound to Sgr~A*, should experience linear drifts on the order of a few $\rm{mas/yr}$. While this uncertainty is certainly relevant for the first few years, proper motion solutions should be well constrained after a few years. More critical is astrometric confusion, which, given the extremely large number of stars in the Galactic Center, will be a limiting factor \citep{Fritz2010}. Due to its transient, stochastic nature, and the unknown stellar distribution and effective AO-performance of MICADO in the Galactic Center, assessing its impact is beyond the scope of this work.

\subsection{Direct detection of SCOs in the Galactic Center}\label{sec:directDetec}

\subsubsection{White dwarfs}
As indicated in \autoref{fig:HR-diagram}, almost all WDs in the GC should be detectable  photometrically with ELT/MICADO, and their detection is facilitated by color measurements. Specifically, their detectability requires precision  color measurements with $\delta \mathrm{H-K} < 0.1$, which has been challenging with VLT data \citep[e.g.,][]{Buchholz2009,Nogueras-Lara2018,Schoedel2020,Gallego-Cano2024}. For instance, while in some cases color-based star classifications have been spectroscopically confirmed, this has only been successful for brighter stars \citep[i.e., $m_K<15\dots 16$;][]{Gallego-Cano2024}. Specifically, while photometric sensitivities are typically on the order $\sim 0.01\dots0.1~\mathrm{mag}$, color measurements are systematically limited by stellar confusion and spatially variable extinction. The ELT will eliminate or at least significantly reduce stellar confusion as source of systematic uncertainty. Still, the spatially variable extinction remains an issue, and it is typically on the order of $\delta A_\lambda \sim 0.2~\mathrm{mag}$ \citep[e.g.,][]{Schodel2010,Fritz2011,Nogueras-Lara2018}. At the same time, the ELT will increase the number of detectable stellar sources by a factor of $10^{3}$–$10^{4}$, thereby enabling substantially improved stellar‑based extinction measurements. The challenge of identifying WDs against the noisy extinction (and confusion) foreground then becomes a well-posed classification problem, where classification models based on our theoretical understanding of stellar evolution are very robust.
It may thus be possible to achieve high enough photometric performance to reliably detect GC WDs, although a dedicated study of this is beyond the scope of this work. 

In addition to normal WDs, $\sim1\%$ of WDs show an infrared excess \citep[e.g.,][]{MadurgaFavieres2024_IRwds}, most likely caused by a warm debris accretion disk.
These systems show significant mid-infrared excess, but their K and H band behavior is much less known. Still, given their typical luminosities of $\sim16~\mathrm{mag}$ in the WISE W1 filter ($\lambda=3.3~\mathrm{\mu m}$) they may form a significant, bright, odd-colored species in the H-R diagram. 

\subsubsection{Stellar black holes}

\begin{figure}  
  \centering
  \includegraphics[width=0.485\textwidth, trim=1cm 1.cm 14cm 4.6cm, clip]{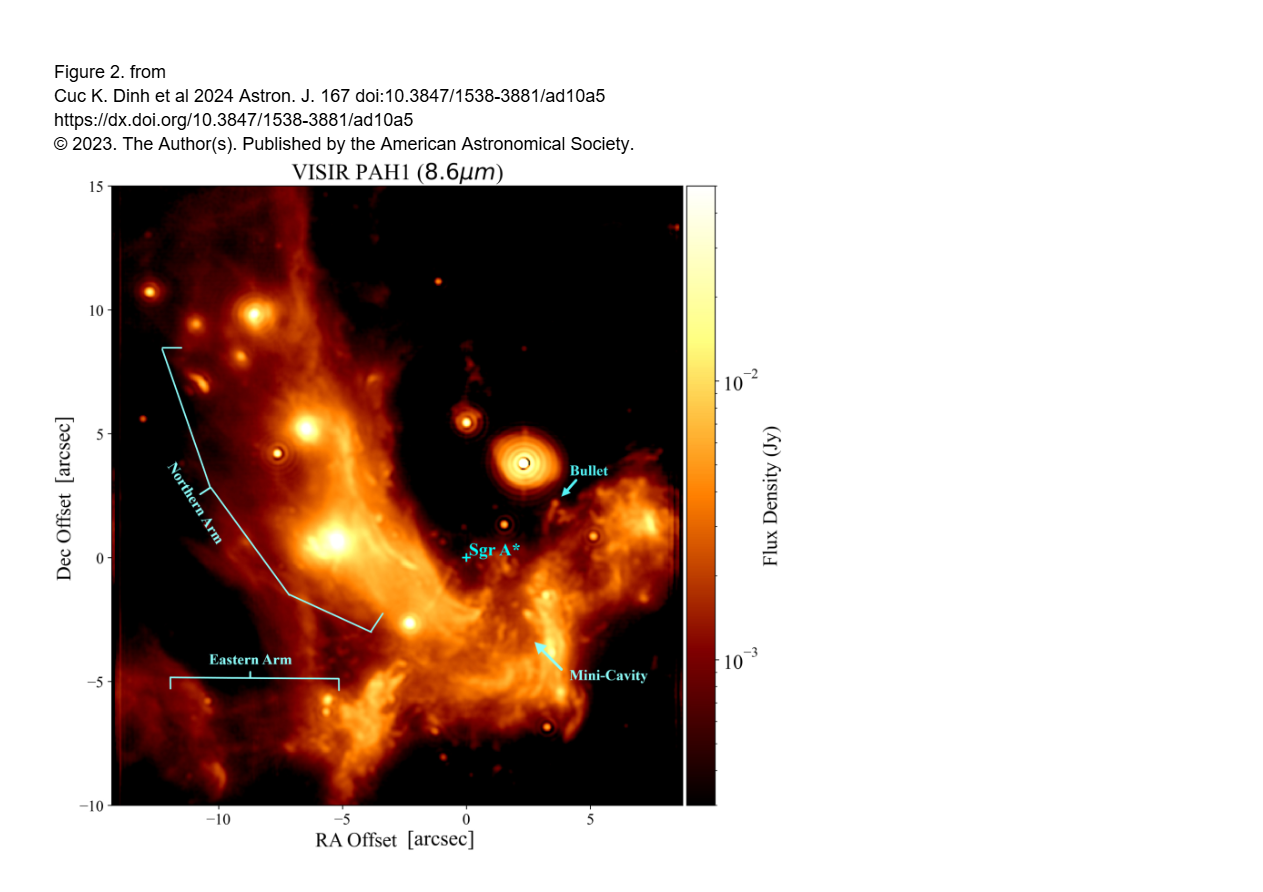}  
  \caption{Paschen-$\alpha$ image of the central $\sim15$ arcseconds$^2$ of the galaxy, showing the large-scale mini-spiral gas streamers. The image was published in \cite{Dinh2024}.}
  \label{fig:MIRimage}
\end{figure}
The Galactic Center hosts an estimated $\approx 10^4$ stellar black holes \citep{Zhang2024} within the central parsec, which appear to be  largely ``dormant'' and have mostly evaded detection. Yet, even isolated BHs may accrete from the gas-rich environment, with accretion states plausibly resembling those of Sgr~A*.
Significant ionized and molecular gas is present (see \autoref{fig:MIRimage}), most prominently in the mini-spiral, a system of gas clouds ($n_{\rm{gas}} = 2 - 21\times 10^4 ~\mathrm{cm^{-3}}$) orbiting around the central MBH \citep[e.g.,][]{Zhao2009, Zhao2010}. Moreover, several tens of Wolf-Rayet stars exist \citep[e.g.,]{Paumard2006,Lu2006,vonFellenberg2022,Jia2023}, whose strong winds likely enhance accretion, and are thought to power the present-day activity of Sgr~A* \citep[e.g.,][]{Quataert2004,Cuadra2008,Shcherbakov2010,Calderon2020b,Ressler2020}.
In this subsection we address the important question: how bright would a stellar black hole be if it accreted from any of those gas sources?

The luminosity of a black hole can be estimated as a fraction of the so-called Eddington luminosity $L_{\rm{Edd}} \propto M_{\rm{BH}}$ [or equivalently, the Eddington accretion rate, $\dot m_{\rm Edd} = L_{\rm Edd}/(\eta c^2$), where $\eta$ is the radiative efficiency and $c$ is the speed of light].
Given Sgr~A*'s luminosity of $\sim 10^{35}\,\mathrm{erg\,s^{-1}}$, a stellar-mass black hole with $M_{\rm BH}=10\,M_{\odot}$ accreting at a comparable fraction of Eddington would plausibly have a luminosity of $L_{\rm SCO}\sim 10^{30}\,\mathrm{erg\,s^{-1}}$, which is roughly an order of magnitude larger than the ELT detection limit in the Galactic Center [$\mathcal{O}(10^{29}~\mathrm{erg/s})$].

\begin{figure*}[htb!]
    \centering
    \includegraphics[width=0.985\textwidth]{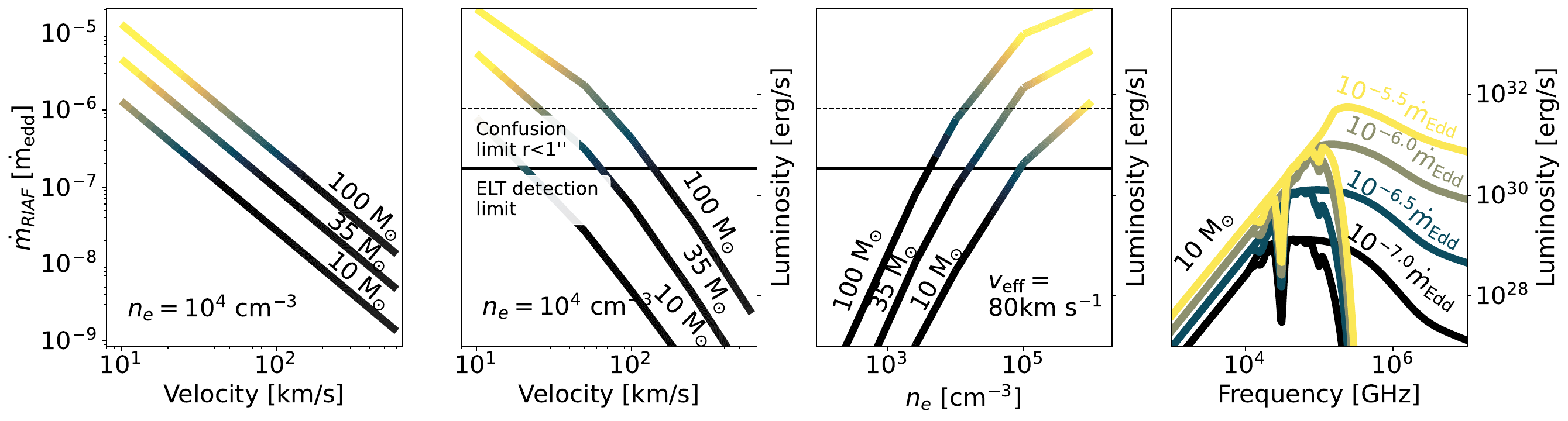}
    \caption{Left panel: Accretion rate in Eddington units for a $10~\mathrm{M_\odot}$ black hole based on scaling relations in \citet{Lalakos2025}, appropriate for radiatively inefficient, turbulent flows. Center two panels: Luminosity of accreting SBHs for different mass ranges, different effective velocity (left center) and different number density of the accreting gas (right center); the color encodes the accretion rate as indicated by the curves in the leftmost panel. The solid horizontal line indicates the ELT detection limit while the dashed horizontal line indicates the confusion and photometric sensitivity multiplied by the extinction factor. Right panel: Model SEDs of a $10~\mathrm{M_{\odot}}$ SBH for different accretion rates computed using the model by \cite{Pesce2021}. 
    The smoother lines do not include the effects of extinction, while the more irregular curves do based on the extinction law by \cite{Fritz2011}.}
    \label{fig:SED}
\end{figure*}

However, relevant for the detectability is the peak frequency at which most of the emission is radiated. Since the emission mechanism for low-luminosity accretion flows is typically synchrotron radiation, this peak frequency is given by the critical frequency of $\nu_b$.
Low accretion rates lead to a so-called Radiatively Inefficient Accretion Flow (RIAF) state, for which $\nu_b \propto M_{\rm{BH}}^{1/2}\cdot \dot{m}^{-1/2}$ \citep[where $M_{\rm{BH}}$ is the SBH mass and $\dot{m}$ is the accretion rate, e.g.,][]{Pesce2021}. This dependence implies that there exists a regime of black hole masses and accretion rates for which the synchrotron emission peaks in the NIR; as illustrated in the right panel \autoref{fig:SED}, this sweet-spot regime is for SBHs with mass ranges between $10\dots 100~\mathrm{M_\odot}$. We have used the \texttt{LLAGN} code by \cite{Pesce2021} to compute the SEDs.

The accretion rate of an SBH that accretes as it moves through the surrounding material can be estimated using the Bondi-Hoyle-Lyttleton accretion formalism \citep[e.g.,][]{HoyleLyttleton1939,BondiHoyle1944,Edgar2004}, where the mass accretion rate is given by \citep[e.g.,][]{Fujita1998}

\begin{equation}
\begin{aligned}
    \dot{m} &\propto \pi r_{\rm cap}^2 \, \rho_{\rm gas} \, V \\
            &\approx 7.4 \times 10^{13} 
            \left( \frac{M}{M_\odot} \right)^2
            \left( \frac{n_{\rm gas}}{10^2 \,\mathrm{cm}^{-3}} \right)
            \left( \frac{v_{\rm eff}}{10 \,\mathrm{km\,s}^{-1}} \right)^{-3}
            \; [\mathrm{g\,s^{-1}}],
\end{aligned}
\label{eqn:AccretionRate}
\end{equation}
where $v_{\rm{eff}}$ is the effective velocity of the gas and the black hole, i.e., $v_{\rm{eff}}=\sqrt{v_{BH}^2 + c_s^2}$, where $v_{\rm{BH}}$ is the relative velocity of the black hole to the surrounding medium and $c_s$ is the sound speed. For typical Galactic Center ionized gas temperatures \citep[$10^4~\mathrm{K}$;][]{Genzel2010} the sound speed is negligible $c_s=11.7 (T/10^4~\mathrm{K})^{1/2} ~\mathrm{km\,s^{-1}} \approx 10 ~\mathrm{km s^{-1}}$ \citep{Wang2014}, setting a lower limit for $v_{\rm{eff}}$, while typical Keplerian velocities are $V \sim \mathcal{O}(10^{2}~\mathrm{km~s}^{-1})$ \citep[e.g.,][]{Genzel2003_stars}. 

However, it is well known that for RIAFs, only a fraction of the gas accreted reaches the event horizon because most of it is lost in the form of outflows. For instance, in the case of Sgr~A*, only $0.01\%$ of the gas accreted at the Bondi radius reaches the black hole event horizon. In general, this result is seen in both simulations and observations of RIAF accretion flows regardless of the physical scale of the system \citep{Xu2019,Gillessen2019}, whether or not the flow is magnetized \citep{Ressler2018,Ressler2020a}, whether or not the flow has significant angular momentum \citep{Pen2003,Pang2011,Ressler2021,Galishnikova2025,Cho2023}, and for several different feeding mechanisms \citep{Ressler2023,Guo2024,Labaj2025}.  
\citet{Xu2023} argue that the significant reduction of the large-scale accretion flow is, in fact, a general property of hot, turbulent accretion, and predict $\dot{m}(r_B)\propto r^{-0.5 \dots 1}$ (consistent with the simulations).   
Recently, \cite{Lalakos2025} performed large scale simulations of Bondi-like accretion onto black holes, for which they claim universality. In these simulations, they find Bondi-radius-to-horizon-scale accretion scaling 
\begin{equation}
    \dot{m}(r_B) = (1.0\pm0.2)\times 10^{-3}\left( \dfrac{r_B}{10^5r_g}\right)^{-0.66\pm 0.03} ~ [\dot{m}_{\rm{Bondi}}],
\end{equation}
where $\dot{m}_{\rm{Bondi}}$ is given as in \autoref{eqn:AccretionRate}, and $r_B$ and $r_g$ is the Bondi and gravitational radius. Here, we have used their relation for Magnetically Arrested Disks (MADs), which is the leading model for Sgr A* \citep[e.g.,][]{Ressler2020,eht_paper_V} and likely a common black hole accretion state for RIAFs \citep{Zamaninasab2014,Chael2019,EHTV_M87,Liska2020}.

The left panel of \autoref{fig:SED} illustrates that SBHs can be detected if they move with and accrete from one of the gas-dense structures like the mini-spiral. Thus, direct constraints on the orbital distribution of SCOs are within reach of the ELT/MICADO. As illustrated in \autoref{fig:HR-diagram}, the actual detection of such SBHs is facilitated by their color, since their synchrotron emission would have a substantially different spectral index than the black-body emitting stars. Important caveats apply to the scaling relations found by \cite{Lalakos2025}, most importantly that they do not explore the impact of different magnetizations of the accreting gas, which is expected to be high in the Galactic Center \citep[e.g.,][]{Eatough2013}.
Additionally, they assume spherical / free-fall accretion on the SBH, that is that the gas does not possess angular momentum, which is likely important for orbiting SBHs. 
Finally, their work does not account for radiative cooling, which would be non-negligible for the initially cold and dense gas accreting from the mini-spiral, even if the flow ultimately remains in the RIAF regime.
Upcoming work by Labaj et al. aims to model such systems in an MHD framework to better constrain the accretion onto SBHs embedded in the mini-spiral environment.

\begin{figure}
    \includegraphics[width=0.485\textwidth]{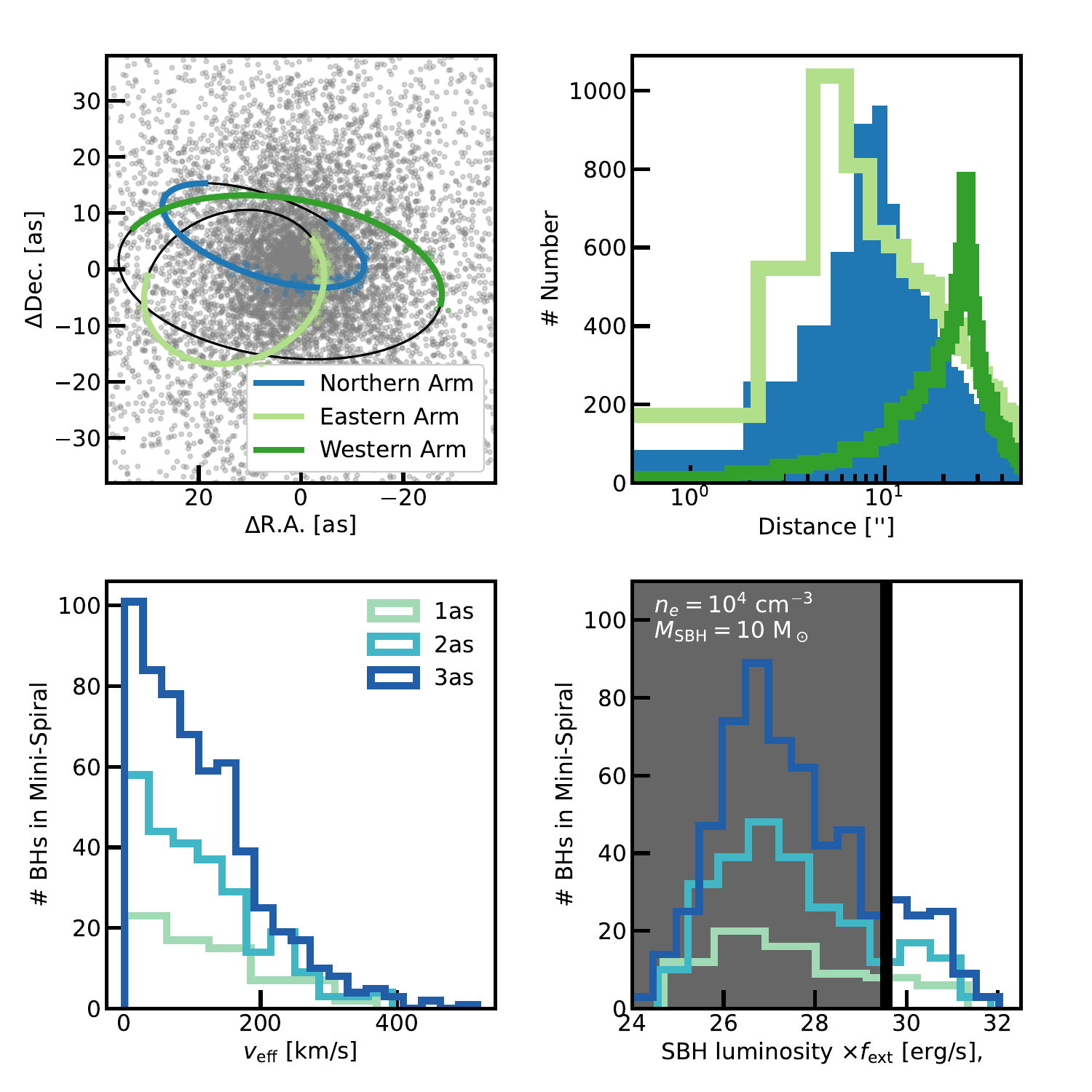}
    \caption{Top left panel: Spatial distribution of SBHs in our model, where SBHs within $1$ arcsecond of the best fit orbits of the three mini-spiral arms are highlighted with the respective colors. 
    Top right panel: Histogram showing the
    distances from the respective best-fit mini-spiral orbits. 
    Bottom left panel: Number of SBHs as a function of effective velocity for three different, assumed mini-spiral widths. 
    Bottom right panel: Respective luminosities of the SBHs, where we have assumed a conservative black hole mass of $10~\mathrm{M_{\odot}}$ and a density of $10^4~\mathrm{cm^{-3}}$ for the accretion scaling laws by \cite{Lalakos2025}; the shaded gray region indicates objects to faint too be detectable.}
    \label{fig:simulatedMiniSpiral}
\end{figure}

To estimate the number of stellar-mass black holes (SBHs) that could be detected with MICADO on the ELT, we first determine their effective velocity relative to the mini-spiral. For the underlying geometry, we adopt the three-dimensional structure of the mini-spiral derived from the orbital elements measured by \cite{Nitschai2020A}. Since the intrinsic width of the mini-spiral is not known, we can only place upper limits based on its projected width of approximately ($2{-}5$ arcsecond, see \autoref{fig:MIRimage}).  We then construct a simulated dark cluster of SBHs, modeled as an isotropic distribution following a power-law density profile \(n_{\rm SBH} \propto r^{1.5}\) \citep{Zhang2024}, with a total population of \(n_{\rm SBH} = 10^{4}\) \citep{GravityCollaboartion2022_massdistribution}. The resulting distributions are illustrated in \autoref{fig:simulatedMiniSpiral}. 

Given our assumptions, we estimate that a few tens of black holes will be above the nominal detection limit -- by up to a factor $10^2$.
Because the spatial distribution of SBHs, the true width of the mini-spiral, the SBH mass function, and the total population remain poorly constrained, this estimate based on our derived luminosity distributions is speculative. 
Still, our assumptions are relatively conservative, as we have chosen a single mass of $10~\mathrm{M_{\odot}}$ for the putative SBHs, and ignored the possibilities that they have experienced mergers or significant accretion during their lifetime. Heavier black holes will be significantly brighter because of the $\dot{m} \propto M_{\rm{BH}}^2$ dependence of accretion. 
Furthermore, our assumption that the mini-spiral is a very small three-dimensional structure of $\sim 1$ arcseconds is highly conservative, as \autoref{fig:MIRimage} seems to indicate that the mini-spiral is significantly larger.
 If we were to use less conservative, more realistic assumptions about the size of the mini-spiral, then we would estimate that $\mathcal{O}(10^2)$ detectable SBHs. 
 Of course, in any case, the most critical assumption of the actual presence of a dark cluster of $\sim 10^4$ SBHs needs to be fulfilled.

\section{Conclusions}
In this paper, we demonstrate that the Galactic Center dark cluster --- the cluster of stellar compact objects bound to Sgr~A* --- can, in principle, be observationally constrained. We demonstrate three different, complementary paths to detection. Photometric detection via stellar ellipsoidal variability, astrometric detection via the orbital astrometric signature of the dark-component in star-SCO binaries, and finally, direct detection of isolated, accreting SBHs. A fourth pathway, via microlensing events of background stars, is photometrically simple, but dedicated surveys are too costly to implement and thus detecting these events can only occur by chance.
Our paper adopts optimistic assumptions about instrument performance, and crucially requires overcoming bright-star confusion in the crowded, luminous, Galactic Center. Most importantly, we do not discuss how to identify these exotic objects in the $\mathcal{O}(10^7)$ stars that reside in the Galactic Center from a practical observational standpoint, which is much more complicated than achieving a high enough signal-to-noise ratio (which we have estimated in this paper). 
This science case is somewhat auxiliary to the obvious case to study the mass, dynamics, and evolutionary state of the Galactic Center with the ELT/MICADO. Given its recently challenged star formation history \citep{Chen2023}, it is clear that a complete and a sufficiently accurate measurement of the stellar properties of the NSC (e.g., \autoref{fig:HR-diagram}) is of central importance to some of the most pressing questions in astronomy, such as what the expected extreme-mass ratio inspiral rate could be for LISA sources \citep[e.g.,][]{LISA_paper2017}, whether or not there could be local foreground gravitational wave signals \citep[e.g.,][]{Xuan2024,AmaroSeoane2025}, and/or how the galaxy and supermassive black hole have coevolved over time \citep[e.g.,][]{Volonteri2010,Chen2023}.
With first closure of the ELT-dome structure, and the completion of the MICADO cryogenic chamber already achieved, this paper demonstrates that the time to prepare Galactic Center observations in the ELT era has come. The extreme amount of data that needs to be processed (by Galactic Center community standards), manipulated, and analyzed, calls for timely preparation. The problem is difficult, but the prize is high.


\begin{acknowledgements}
    We thank our referee, Mark Morris, for a thorough review and comments that substantially contributed to the manuscript. SDvF thanks Mark Morris and Gunther Witzel for the fruitful discussions on the Galactic Center dark cluster.
    SDvF gratefully acknowledges the support of the Alexander von Humboldt Foundation through a Feodor Lynen Fellowship and thanks CITA for their hospitality and collaboration; and is grateful for scientific input on the manuscript to Reinhard Genzel, Frank Eisenhauer, Stefan Gillessen, and all members of MPE GC team. We thank Michal Zaja{\v c}ek and Aris Lalakos for the discussion on radiatively inefficient accretion onto compact objects. We thank Mark Dodici for the insightful discussions on Galactic Center binary star properties; and Aleksey Generozov on the help implementing and understanding the stellar evolutionary track models.
    This manuscript was prepared with the assistance of an AI-based large language model (Microsoft Copilot, GPT-5 family) for language editing and Python code generation. The authors take full responsibility for the scientific content, data analysis, and interpretation presented in this work. All code was reviewed, validated, and executed by the authors.
    SDvF and SMR acknowledge the support of the Natural Sciences and Engineering Research Council of Canada (NSERC), [funding reference number 568580] Cette recherche a \'et\'e financ\'ee par le Conseil de recherches en sciences naturelles et en g\'enie du Canada (CRSNG), [num\'ero de r\'ef\'erence 568580]. 
    SDvF is supported by the Canadian Space Agency (23JWGO2A01). 
    ML acknowledges the GACR Junior Star grant no. GM24-10599M for support.
\end{acknowledgements}

%
\bibliographystyle{aa} 
\bibliography{references} 

\begin{appendix}





\end{appendix}
\end{document}